\RequirePackage{lineno}

\documentclass[aps,prc,twocolumn,groupedaddress,showpacs,amsmath,amssymb,floatfix,superscriptaddress]{revtex4}
\usepackage{multirow}
\usepackage{graphicx}
\usepackage{times}
\usepackage{color}

\begin{document}

\bibliographystyle{h-physrev3}

\title{Limits on Majoron-emitting double-beta decays of $^{136}$Xe in the KamLAND-Zen experiment}

\newcommand{\tohoku}{\affiliation{Research Center for Neutrino
    Science, Tohoku University, Sendai 980-8578, Japan}}
\newcommand{\osaka}{\affiliation{Graduate School of 
    Science, Osaka University, Toyonaka, Osaka 560-0043, Japan}}
\newcommand{\lbl}{\affiliation{Physics Department, University of
    California, Berkeley, and \\ Lawrence Berkeley National Laboratory, 
Berkeley, California 94720, USA}}
\newcommand{\colostate}{\affiliation{Department of Physics, Colorado
    State University, Fort Collins, Colorado 80523, USA}}
\newcommand{\ut}{\affiliation{Department of Physics and
    Astronomy, University of Tennessee, Knoxville, Tennessee 37996, USA}}
\newcommand{\tunl}{\affiliation{Triangle Universities Nuclear
    Laboratory, Durham, North Carolina 27708, USA and \\
Physics Departments at Duke University, North Carolina Central University,
and the University of North Carolina at Chapel Hill}}
\newcommand{\ipmu}{\affiliation{Kavli Institute for the Physics and Mathematics of the Universe (WPI),
University of Tokyo, Kashiwa, 277-8583, Japan}}
\newcommand{\nikhef}{\affiliation{Nikhef and the University of Amsterdam, Science Park, Amsterdam, the Netherlands}}
\newcommand{\washington}{\affiliation{Center for Experimental Nuclear Physics and Astrophysics, University of Washington, Seattle, Washington 98195, USA}}

%
%
\author{A.~Gando}\tohoku
\author{Y.~Gando}\tohoku
\author{H.~Hanakago}\tohoku
\author{H.~Ikeda}\tohoku
\author{K.~Inoue}\tohoku\ipmu
\author{R.~Kato}\tohoku
\author{M.~Koga}\tohoku\ipmu
\author{S.~Matsuda}\tohoku
\author{T.~Mitsui}\tohoku
\author{T.~Nakada}\tohoku
\author{K.~Nakamura}\tohoku\ipmu
\author{A.~Obata}\tohoku
\author{A.~Oki}\tohoku
\author{Y.~Ono}\tohoku
\author{I.~Shimizu}\tohoku
\author{J.~Shirai}\tohoku
\author{A.~Suzuki}\tohoku
\author{Y.~Takemoto}\tohoku
\author{K.~Tamae}\tohoku
\author{K.~Ueshima}\tohoku
\author{H.~Watanabe}\tohoku
\author{B.D.~Xu}\tohoku
\author{S.~Yamada}\tohoku
\author{H.~Yoshida}\tohoku

\author{A.~Kozlov}\ipmu

\author{S.~Yoshida}\osaka

\author{T.I.~Banks}\lbl
\author{J.A.~Detwiler}\lbl
\author{S.J.~Freedman}\ipmu\lbl
\author{B.K.~Fujikawa}\ipmu\lbl
\author{K.~Han}\lbl
\author{T.~O'Donnell}\lbl

\author{B.E.~Berger}\colostate

\author{Y.~Efremenko}\ipmu\ut

\author{H.J.~Karwowski}\tunl
\author{D.M.~Markoff}\tunl
\author{W.~Tornow}\tunl

\author{S.~Enomoto}\ipmu\washington

\author{M.P.~Decowski}\ipmu\nikhef

\collaboration{KamLAND-Zen Collaboration}\noaffiliation

\date{\today}

\begin{abstract}
We present limits on Majoron-emitting neutrinoless double-beta decay modes based on an exposure of 112.3~days with 125~kg of $^{136}$Xe. In particular, a lower limit on the ordinary (spectral index $n = 1$) Majoron-emitting decay half-life of $^{136}$Xe is obtained as $T_{1/2}^{0\nu\chi^{0}} > 2.6 \times 10^{24}$\,yr at 90\% C.L., a factor of five more stringent than previous limits. The corresponding upper limit on the effective Majoron-neutrino coupling, using a range of available nuclear matrix calculations, is $\left<g_{ee}\right> < (0.8 - 1.6) \times 10^{-5}$. This excludes a previously unconstrained region of parameter space and strongly limits the possible contribution of ordinary Majoron emission modes to $0\nu\beta\beta$ decay for neutrino masses in the inverted hierarchy scheme.
\end{abstract}

\pacs{23.40.$-$s, 21.10.Tg, 14.60.Pq, 27.60.$+$j}

\maketitle

The search for neutrinoless double-beta ($0\nu\beta\beta$) decay is the best probe of the Majorana nature of the neutrino known at present. The observation of this process would immediately imply total lepton number violation and the equivalence of the neutrino and the antineutrino, irrespective of the mechanism by which the decay is mediated~\cite{Schechter1982}. Although most current experimental efforts focus on the detection of $0\nu\beta\beta$ decay mediated by light Majorana neutrino exchange, many other mechanisms have been proposed. Some exotic models~\cite{Gelmini1981,Doi1985} predict decays proceeding through the emission of massless Nambu-Goldstone (NG) bosons, referred to as Majorons. Precise measurements of the invisible decay width of the Z boson in LEP~\cite{LEP2006} showed that traditional Majoron models require severe fine-tuning~\cite{Gunther1996}. However, a number of additional models have been proposed which avoid such fine tuning, including modes in which the Majoron can carry leptonic charge, and need not be a NG boson~\cite{Burgess1994}, or in which $0\nu\beta\beta$ decay proceeds through the emission of two Majorons~\cite{Bamert1995}. These models predict different shapes for the spectrum of the summed energy of the two emitted $\beta$'s. In this report we analyze the spectrum obtained from a 38.6\,kg-year exposure of $^{136}$Xe with KamLAND-Zen~\cite{Gando2012} to derive new limits on several of these decay modes.

\begin{table*}
\caption{\label{table:limit}
Different Majoron emission models~\cite{Bamert1995,Carone1993,Mohapatra2000,Hirsch1996}, and the KamLAND-Zen limits for the corresponding Majoron-emitting $0\nu\beta\beta$ decay half-lives ($T_{1/2}$) and effective Majoron-neutrino coupling constants ($\left<g_{ee}\right>$) for $^{136}$Xe at 90\% C.L. The model notation in the first column follows Refs.~\cite{Bamert1995} and \cite{Arnold2006}. 
The third, fourth, fifth, and sixth columns indicate whether the Majoron is a NG boson or not, its leptonic charge ($L$), the model's spectral index ($n$), and the form of the nuclear matrix element, respectively.}
\begin{ruledtabular}
\begin{tabular}{llllllll}
\multirow{2}{*}{Model} & \multirow{2}{*}{Decay mode} & \multirow{2}{*}{NG boson} & \multirow{2}{*}{$L$} & \multirow{2}{*}{$n$} & \multirow{2}{*}{Matrix element} & \multicolumn{2}{l}{Results from this measurement}   \\
 & & & & & & $T_{1/2}$\,(yr) & $\left<g_{ee}\right>$ \\
\hline
IB & $0\nu\beta\beta\chi^{0}$ & No & 0 & 1 & $M_{F} - M_{GT}$~\cite{Simkovic2009,Menendez2009} & $> 2.6 \times 10^{24}$ & $< (0.8 - 1.6) \times 10^{-5}$ \\
IC & $0\nu\beta\beta\chi^{0}$ & Yes & 0 & 1 & $M_{F} - M_{GT}$~\cite{Simkovic2009,Menendez2009} & $> 2.6 \times 10^{24}$ & $< (0.8 - 1.6) \times 10^{-5}$ \\
ID & $0\nu\beta\beta\chi^{0}\chi^{0}$ & No & 0 & 3 & $M_{F\omega^{2}} - M_{GT\omega^{2}}$~\cite{Hirsch1996} & $> 4.5 \times 10^{23}$ & $< 0.68$ \\
IE & $0\nu\beta\beta\chi^{0}\chi^{0}$ & Yes & 0 & 3 & $M_{F\omega^{2}} - M_{GT\omega^{2}}$~\cite{Hirsch1996} & $> 4.5 \times 10^{23}$ & $< 0.68$ \\
IIB & $0\nu\beta\beta\chi^{0}$ & No & $-2$ & 1 & $M_{F} - M_{GT}$~\cite{Simkovic2009,Menendez2009} & $> 2.6 \times 10^{24}$ & $< (0.8 - 1.6) \times 10^{-5}$ \\
IIC & $0\nu\beta\beta\chi^{0}$ & Yes & $-2$ & 3 & $M_{CR}$~\cite{Hirsch1996} & $> 4.5 \times 10^{23}$ & $< 0.013$ \\
IID & $0\nu\beta\beta\chi^{0}\chi^{0}$ & No & $-1$ & 3 & $M_{F\omega^{2}} - M_{GT\omega^{2}}$~\cite{Hirsch1996} & $> 4.5 \times 10^{23}$ & $< 0.68$ \\
IIE & $0\nu\beta\beta\chi^{0}\chi^{0}$ & Yes & $-1$ & 7 & $M_{F\omega^{2}} - M_{GT\omega^{2}}$~\cite{Hirsch1996} & $> 1.1 \times 10^{22}$ & $< 1.2$ \\
IIF & $0\nu\beta\beta\chi^{0}$ & Gauge boson & $-2$ & 3 & $M_{CR}$~\cite{Hirsch1996} & $> 4.5 \times 10^{23}$ & $< 0.013$ \\
\vspace{-3mm} \\
``bulk'' & $0\nu\beta\beta\chi^{0}$ & Bulk field & 0 & 2 & -- & $> 1.0 \times 10^{24}$ & -- \\
\end{tabular}
\end{ruledtabular}
\end{table*}

Table~\ref{table:limit} summarizes ten Majoron models~\cite{Bamert1995,Carone1993,Mohapatra2000,Hirsch1996}, which can be divided into two categories: (I) lepton number violating models, and (II) lepton number conserving models. The table indicates also whether the corresponding $0\nu\beta\beta$ decay is accompanied by the emission of one or two Majorons: 
\begin{eqnarray}
(A, Z) & \rightarrow & (A, Z + 2) + 2e^{-} + \chi^{0}\;, \\
(A, Z) & \rightarrow & (A, Z + 2) + 2e^{-} + 2\chi^{0}\;.
\end{eqnarray}
The main distinguishing features of the models are listed in the third, fourth, and fifth columns: whether the Majoron is a NG boson or not, its leptonic charge ($L$), and the model's spectral index ($n$), respectively. The spectral index is defined from the phase space of the emitted particles, $G \sim (Q_{\beta\beta} - K)^{n}$, where $Q_{\beta\beta}$ is the $Q$ value of the $\beta\beta$ decay and $K$ is the summed energy of the electrons. A spectral index of $n = 1$ is denoted as ordinary Majoron emission, while $n \neq 1$ is referred to as non-ordinary. Experimental searches for $\beta\beta$ decay mediated by Majoron emission have been performed by Heidelberg-Moscow for $^{76}$Ge~\cite{Gunther1996}; by NEMO-2 and NEMO-3 for $^{100}$Mo, $^{116}$Cd, $^{82}$Se, $^{96}$Zr, $^{150}$Nd, and $^{130}$Te~\cite{Arnold2000, Arnold2006, Argyriades2009, Argyriades2010, Arnold2011, Barabash2011b}; by ELEGANT V for $^{100}$Mo~\cite{Fushimi2002}; and by DAMA for $^{136}$Xe~\cite{Bernabei2002}. Each of these searches reported limits for ordinary Majoron emission, while several also reported limits for non-ordinary emission modes.

\begin{center}
\begin{table}[b]
\caption{\label{table:systematic}Estimated systematic uncertainties used for the $^{136}$Xe $\beta\beta$ decay half-life measurement. The overall uncertainty is 5.2\%.
}
\begin{tabular}{@{}*{2}{lc}}
\hline
\hline
Source  \hspace{4.0cm} & Systematic Uncertainty (\%) \\
\hline
Fiducial volume & 5.2 \\
Enrichment factor of $^{136}$Xe & 0.05 \\
Xenon concentration & 0.34 \\
Detector energy scale & 0.3 \\
Xe-LS edge effect & 0.06 \\
Detection efficiency & 0.2 \\
\hline
Total & 5.2 \\
\hline
\hline
\end{tabular}
\end{table}
\end{center}

The experimental investigation reported here is based on data collected with KamLAND-Zen between October 12, 2011, and February 9, 2012, and includes the data used in Ref.~\cite{Gando2012} plus an additional 34.7\,days, for a total livetime of 112.3\,days. The target liquid scintillator (LS) contains $(2.44 \pm 0.01)\%$ by weight of enriched xenon gas, obtained by augmenting the measurement by gas chromatography, $(2.52 \pm 0.07)\%$~\cite{Gando2012}, with a much more precise estimate based on the total xenon weight and LS volume introduced into the detector during LS filling, $(2.44 \pm 0.01)\%$. Reconstructed energies and positions of muon-induced neutron-capture $\gamma$ events confirm that the detector status and data quality were stable over the data-set. The fiducial volume for the reconstructed event vertices is defined as a spherical shape 1.2~m in radius at the detector center, containing $(125 \pm 7)\,{\rm kg}$ of $^{136}$Xe. Other event selection criteria are applied as described in Ref.~\cite{Gando2012}. The systematic uncertainties are summarized in Table~\ref{table:systematic}; the dominant contribution comes from the fiducial volume uncertainty. The total systematic uncertainty on the $\beta\beta$ decay half-life measurement is 5.2\%, slightly improved from the 5.9\% of Ref.~\cite{Gando2012} owing to the supplemental xenon concentration measurement.

\begin{figure}
\begin{center}
\vspace{0.2cm}
\includegraphics[angle=270,width=\columnwidth]{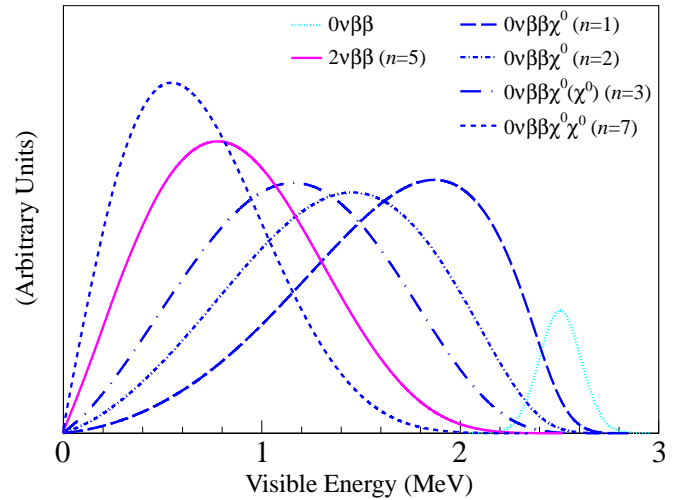}
\vspace{-0.35cm}
\end{center}
\caption[]{KamLAND-Zen visible energy spectra for different $^{136}$Xe decay modes, characterized by the spectral index $n$. The resolution-limited line [$\sigma$ = ($6.6 \pm 0.3)\%/\sqrt{E({\rm MeV})}$] at the \mbox{$Q$ value} indicates the $0\nu\beta\beta$ decay peak without Majoron emission.}
\label{figure:model}
\end{figure}

Figure~\ref{figure:model} shows predicted energy spectra of $^{136}$Xe decay for different values of the spectral index $n$, corresponding to $2\nu\beta\beta$ ($n = 5$), $0\nu\beta\beta\chi^{0}$ ($n = 1,2,$ and $3$), and $0\nu\beta\beta\chi^{0}\chi^{0}$ ($n = 3$ and $7$). The spectra have been convolved with the KamLAND-Zen detector response function, including the energy resolution and energy scale non-linearities. In this analysis, the dominant contribution from $2\nu\beta\beta$ decay is fit simultaneously with a possible contribution from one of the $0\nu\beta\beta$ modes to obtain an upper limit on its rate.

\begin{figure}
\begin{center}
\includegraphics[angle=0,width=1.0\columnwidth]{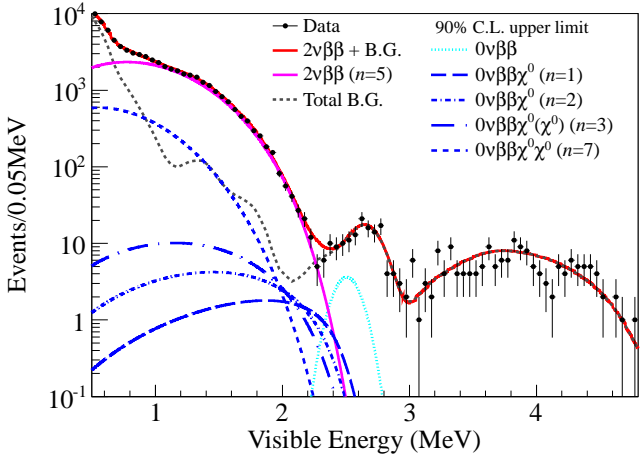}
\vspace{-1.0cm}
\end{center}
\caption[]{Energy spectrum of selected $\beta\beta$ decay candidates (data points) together with the best-fit backgrounds (gray dashed line) and $2\nu\beta\beta$ decay (purple solid line), and the 90\% C.L. upper limit for $0\nu\beta\beta$ decay and Majoron-emitting $0\nu\beta\beta$ decays for each spectral index. The red line depicts the sum of the $2\nu\beta\beta$ decay and background spectra. Numerical results are reported in Table~\ref{table:limit}. The best-fit has a $\chi^{2}$/d.o.f. = 100.4/87 for the full fit range $0.5 < E < 4.8~{\rm MeV}$.}
\label{figure:energy}
\end{figure}

\begin{figure}
\begin{center}
\vspace{0.0cm}
\includegraphics[angle=0,width=1.0\columnwidth]{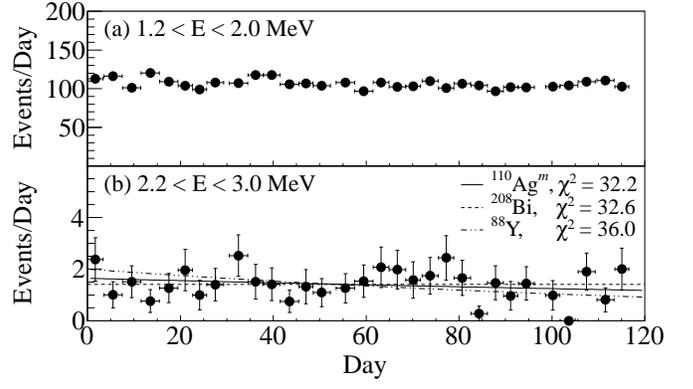}
\vspace{-0.7cm}
\end{center}
\caption[]{Event rate variation in the energy regions (a) $1.2 < E < 2.0~{\rm MeV}$ ($2\nu\beta\beta$ window) and (b) $2.2 < E < 3.0~{\rm MeV}$ ($0\nu\beta\beta$ window). The fitted curves correspond to the expected variations for the hypotheses that all the events in the $0\nu\beta\beta$ window are solely from one of the background candidates, $^{\rm 110}$Ag$^{m}$ (solid line), $^{208}$Bi (dotted line), or unsupported $^{88}$Y (dot-dashed line).}
\label{figure:time}
\end{figure}

The energy spectrum of selected candidate events between 0.5 and 4.8 MeV is shown in Fig.~\ref{figure:energy}. The $\beta\beta$ decay rates are estimated from a likelihood fit to the binned energy spectrum. In the fit, background contributions from external sources, from the $^{222}$Rn-$^{210}$Pb and $^{228}$Th-$^{208}$Pb chains, and from $^{11}$C and $^{10}$C (muon spallation products), as well as the parameters of the detector energy response model are allowed to vary but are constrained by their independent measurement~\cite{Gando2012}. Backgrounds without independent measurements, namely $^{85}$Kr, $^{40}$K, non-equilibrium $^{210}$Bi, and the $^{238}$U-$^{222}$Rn and $^{232}$Th-$^{228}$Th decay chains, are left unconstrained. As in Ref.~\cite{Gando2012}, the background in the $0\nu\beta\beta$ region of interest, $2.2 < E < 3.0~{\rm MeV}$, is fit to a combination of $^{110}$Ag$^{m}$, $^{88}$Y, $^{208}$Bi, and $^{60}$Co, constrained by the observed time variation of the event rate in that energy range, shown in Fig.~\ref{figure:time}.

The additional exposure in this analysis yields an improved measurement of the $2\nu\beta\beta$ decay half-life of $^{136}$Xe. Setting the contributions from all Majoron-emitting modes to zero gives $T_{1/2}^{2\nu} = 2.30 \pm 0.02({\rm stat}) \pm 0.12({\rm syst}) \times 10^{21}$~yr, which is consistent with the previous result~\cite{Gando2012}. The half-life limit for $0\nu\beta\beta$ decay also improves slightly to $T_{1/2}^{0\nu} > 6.2 \times 10^{24}$~yr at 90\% C.L.

The event rate in the $0\nu\beta\beta$ region of interest shows little time variation, limiting the allowed contribution of $^{88}$Y (\mbox{$T_{1/2}=107$~days}) assuming its parent $^{88}$Zr is absent. There is little statistical power to distinguish $^{110}$Ag$^{m}$ (\mbox{$T_{1/2}=250$~days}) and $^{208}$Bi (\mbox{$T_{1/2}=3.7 \times 10^5$~yr}). However, the energy spectrum without $^{\rm 110}$Ag$^{m}$ is rejected by a $\chi^{2}$ test at more than $3\sigma$ C.L., indicating a preference for a dominant contribution from this contaminant in the $0\nu\beta\beta$ region. We find that including a non-zero contamination of $^{88}$Zr changes only the relative contributions of $^{88}$Y and $^{208}$Bi, but there is no impact on the other spectral components.

The 90\% C.L. upper limits for the different Majoron-emitting decay mode spectra are drawn in Fig.~\ref{figure:energy}, and the corresponding half-life limits are listed in Table~\ref{table:limit}. In particular, the lower limit on the ordinary (spectral index $n = 1$) Majoron-emitting decay half-life is $T_{1/2}^{0\nu\chi^{0}} > 2.6 \times 10^{24}$\,yr at 90\% C.L., which is a factor of five more stringent than previous limits~\cite{Bernabei2002}. Owing to the larger background at lower energy, the sensitivity to higher spectral index decays is weaker.

The limits on single- or double-Majoron emission can be translated into limits on the effective coupling constant of the Majoron to the neutrino, $\left<g_{ee}\right>$~\cite{Doi1985}, using the relations
\begin{eqnarray}
T_{1/2}^{-1} & = & \left| \left<g_{ee}\right> \right|^{2} \left| M \right|^{2} G ~~~ \mbox{for $0\nu\beta\beta\chi^{0}$}\;, \\
T_{1/2}^{-1} & = & \left| \left<g_{ee}\right> \right|^{4} \left| M \right|^{2} G ~~~ \mbox{for $0\nu\beta\beta\chi^{0}\chi^{0}$}\;.
\end{eqnarray}
The nuclear matrix elements $M$ and the phase space factors $G$ for the ordinary Majoron-emitting decay ($n = 1$) are taken from~\cite{Simkovic2009, Menendez2009} and~\cite{Suhonen1998}, respectively, while those for other decays are taken from~\cite{Hirsch1996}. From the half-life limit of the ordinary Majoron-emitting decay mode we obtain an upper limit of \mbox{$\left<g_{ee}\right> < (0.8 - 1.6) \times 10^{-5}$} at 90\% C.L., where the range of the upper limit corresponds to the theoretical range of the nuclear matrix elements~\cite{Simkovic2009, Menendez2009}. This is the most stringent limit on $\left<g_{ee}\right>$ to date among all $\beta\beta$ decay nuclei~\cite{Gunther1996,Arnold2000, Arnold2006, Argyriades2009, Argyriades2010, Arnold2011,Barabash2011b,Fushimi2002,Bernabei2002}. The previous best limit from a laboratory experiment was from NEMO-3 for $^{100}$Mo: \mbox{$\left<g_{ee}\right> <  (3.5 - 8.5) \times 10^{-5}$}~\cite{Barabash2011b}. Our new limit corresponds to more than a factor of 2.2 improvement over this previous result. The limits on the effective Majoron-neutrino coupling constant for $^{136}$Xe for all investigated Majoron-emitting $0\nu\beta\beta$ decays are summarized in Table~\ref{table:limit}.

Other limits on $\left<g_{ee}\right>$ are available from geochemistry and astrophysics. Half-life limits on $^{128}$Te from geochemical experiments can be interpreted as an effective coupling limit of \mbox{$\left<g_{ee}\right> < 3 \times 10^{-5}$}~\cite{Bernatowicz1992}, although the half-life determinations have been criticized~\cite{Thomas2008} and may require a downward correction by almost a factor of 3. The observation of neutrinos from SN1987A and of their time distribution indicates that Majoron emission does not play a dominant role in core collapse processes, allowing one to exclude the range \mbox{$4 \times 10^{-7} < \left<g_{ee}\right> < 2 \times 10^{-5}$}~\cite{Kachelriess2000,Tomas2001,Farzan2003} for the ordinary Majoron-emitting decay mode. While previous limits combined with the supernova data still allowed a gap region of \mbox{$2 \times 10^{-5} < \left<g_{ee}\right> < 9 \times 10^{-5}$}~\cite{Barabash2011b}, our new result completely excludes this region. The SN1987A limit therefore significantly extends the KamLAND-Zen limit down to \mbox{$\left<g_{ee}\right> < 4 \times 10^{-7}$}. Multiplying by the square-root of the ratio of phase space factors, one finds that this limit excludes the possibility that ordinary Majoron-emitting decay modes play a dominant role to light Majorana neutrino exchange for $\left<m_{\beta\beta}\right> > 20\,{\rm meV}$. This range covers almost the entire $\left<m_{\beta\beta}\right>$ parameter space in the case of the inverted neutrino mass hierarchy~\cite{Nakamura2010}.

In summary, we have reported new limits from KamLAND-Zen on Majoron-emitting $0\nu\beta\beta$ decay half-lives for $^{136}$Xe. In particular, for ordinary Majoron-emitting $0\nu\beta\beta$ decay (spectral index $n = 1$), we obtained an improved limit on the effective Majoron-neutrino coupling constant $\left<g_{ee}\right>$ by more than a factor of 2.2 over earlier searches. Combined with limits on $\left<g_{ee}\right>$ from SN1987A, the KamLAND-Zen result strongly disfavors a dominant contribution from ordinary Majoron-emitting decay modes for neutrino masses in the inverted hierarchy scheme.

The KamLAND-Zen experiment is supported by the Grant-in-Aid for Specially Promoted Research under grant 21000001 of the Japanese Ministry of Education, Culture, Sports, Science and Technology; the World Premier International Research Center Initiative (WPI Initiative), MEXT, Japan; and under the US Department of Energy (DOE) Grant No. DE-AC02-05CH11231, as well as other DOE grants to individual institutions. The Kamioka Mining and Smelting Company has provided service for activities in the mine.

\bibliography{DoubleBetaMajoron}

\end{document}